\documentclass[journal=langd5,manuscript=article]{achemso}
\usepackage{amsmath}
\usepackage{amssymb}
\usepackage{graphicx}
\usepackage{dcolumn}
\usepackage{bm}
\usepackage{color}

\usepackage[colorlinks=true,linkcolor=magenta,citecolor=cyan]{hyperref}
\title{Enhanced premelting at the ice--rubber interface using all-atom molecular dynamics simulation}

\author{Takumi Kojima}
\affiliation{Department of Mechanical Engineering, Keio University, Yokohama, Kanagawa 223-8522, Japan}
\author{Ikki Yasuda}
\affiliation{Department of Mechanical Engineering, Keio University, Yokohama, Kanagawa 223-8522, Japan}
\author{Takumi Sato}
\affiliation{Department of Mechanical Engineering, Keio University, Yokohama, Kanagawa 223-8522, Japan}
\author{Noriyoshi Arai}   
\affiliation{Department of Mechanical Engineering, Keio University, Yokohama, Kanagawa 223-8522, Japan}
\author{Kenji Yasuoka}   
\email{yasuoka@mech.keio.ac.jp}
\affiliation{Department of Mechanical Engineering, Keio University, Yokohama, Kanagawa 223-8522, Japan}

\begin{document}
\begin{abstract}
The ice–rubber interface is critical in applications such as tires and shoe outsoles, yet its molecular tribology remains unclear. Using all-atom molecular dynamics simulations, we studied premelting layers at the basal face of ice in contact with styrene–butadiene rubber from 254 to 269 K. Despite its hydrophobicity, rubber enhances structural disorder of interfacial water, promoting premelting. In contrast, water mobility is suppressed by confinement from polymer chains, leading to glassy dynamics distinct from the ice–vapor interface. Near the melting point, rubber chains become more flexible and penetrate the premelting layer, forming a mixed rubber–water region that couples the dynamics of both components. These results suggest that nanoscale roughness and morphology of hydrophobic polymers disrupt ice hydrogen-bond networks, thereby enhancing premelting. Our findings provide molecular-level insight into ice slipperiness and inform the design of polymer materials with controlled ice adhesion and friction.
\end{abstract} 

\section{Introduction}
The ice–rubber interface plays a key role in many industrial applications, including vehicle tires and shoe outsoles. For tires, rubber materials help reduce the risk of slipping and allow efficient transmission of engine power and braking force to icy road surfaces. Rubber friction arises from its viscoelasticity and surface adhesion on solid surfaces, typically showing a bell-shaped dependence on sliding velocity~\cite{grosch1963relation, ROBERTS197675, chernyak1986theory}. In contrast, the tribology of rubber on ice differs markedly because of the unique properties of ice. Ice is exceptionally slippery, and its low friction has been attributed to several mechanisms, including frictional heating, premelting layers, pressure melting, and the formation of thin liquid-like surface films~\cite{southern1972friction, persson2015ice, rosenberg2005ice, kietzig2010physics,tuononen2016multiscale,hemette2021thermal}. 
\textcolor{black}{In recent years, systematic experiments have revealed that the interfacial temperature affects the thickness of the viscoelastic ice surface film~\cite{canale2019nanorheology}, that the friction mode changes depending on surface temperature and pressure~\cite{liefferink2021friction}, and that surface chemistry modulates molecular diffusion and, consequently, friction~\cite{canale2019nanorheology, baran2022ice}.} 
\textcolor{black}{Despite these advances in understanding ice tribology, the microscopic origins of ice–rubber interfacial tribology remain poorly understood.}

A thin liquid-like layer, known as the premelting layer, forms on ice surfaces and strongly affects interfacial behavior~\cite{dash2006physics, li2007surface, slater2019surface}. Its thickness has been reported from a few nanometers to micrometers, and although debated, it is widely accepted that the layer behaves like a liquid. 
Near the melting point, it is highly sensitive to temperature, where changes of only a few kelvins can strongly alter its morphology, structure, and dynamics~\cite{sazaki2012quasi, sanchez2017experimental, nagata2019surface, llombart2020surface}.
Molecular simulations have been widely used to probe the structure and dynamics of premelting layers, supporting their liquid-like nature~\cite{conde2008thickness, pfalzgraff2011comparative, sanchez2017experimental, kling2018structure, pickering2018grand, qiu2018so, llombart2020surface}. 
Recent computational studies have shown that small molecules can locally alter ice surfaces. For example, ions disrupt the local ice structure~\cite{hudait2017sink, berrens2022effect}, and \textcolor{black}{hydrophilic polymers of low molecular weight induce partial melting and suppress ice growth~\cite{bachtiger2021atomistic}}. 
These computational studies have provided molecular insights into ice–vapor interfaces.

The ice–polymer (or rubber) interface has been studied experimentally for decades, demonstrating that the surface properties of ice are strongly influenced by the chemistry and microstructure of the contacting material.
Friction measurements showed that the viscosity of the premelting film is sensitive to surface chemistry~\cite{canale2019nanorheology}. 
Hydrophobic coatings reduce ice adhesion, whereas hydrophilic polymers interact with liquid-like films to form hydration layers~\cite{chen2017icephobic, pallbo2024enhanced}.
Lecadre et al.\cite{lecadre2020ice} suggested that friction is governed by the viscoelasticity of contacting materials at low temperatures, but dominated by premelting layers at higher temperatures. 
Molecular dynamics simulations further revealed that wettability strongly affects premelting-layer friction\cite{baran2022ice}, with similar behavior seen in supercooled water~\cite{gasparotto2022interfaces}. More realistic interface models, including polymer and graphene interfaces, showed that hydrogen bonding with the contacting surface alters the orientation, structure, and dynamics of water molecules in the premelting layers~\cite{skountzos2024interfacial}.
These findings suggest that both the chemical composition and nanostructure of rubber chains can modify premelting layers, producing behaviors distinct from those at the pure ice–vapor interface. However, the atomistic details of such rubber-induced effects remain largely unexplored.

Rubber is mainly composed of nonpolar polymers such as styrene and butadiene, often combined with cross-linkers and fillers. Its hydrophobic nature is expected to affect the ice surface through Van der Waals interactions.
In this study, we use molecular dynamics simulations to investigate the atomistic structure and dynamics at the ice–rubber interface, as slipperiness is closely linked to molecular mobility~\cite{weber2018molecular}.
We find that rubber enhances structural disorder in the premelting layers while suppressing their dynamics due to confinement.
Importantly, we also show that the dynamics of interfacial rubber chains change cooperatively with premelting water molecules, and at near the melting temperature, a mixed rubber–water layer emerges.

\section{Materials and Methods}
\subsection{Molecular Dynamics Simulation}
\begin{figure}[t!]
\includegraphics[width=90mm]{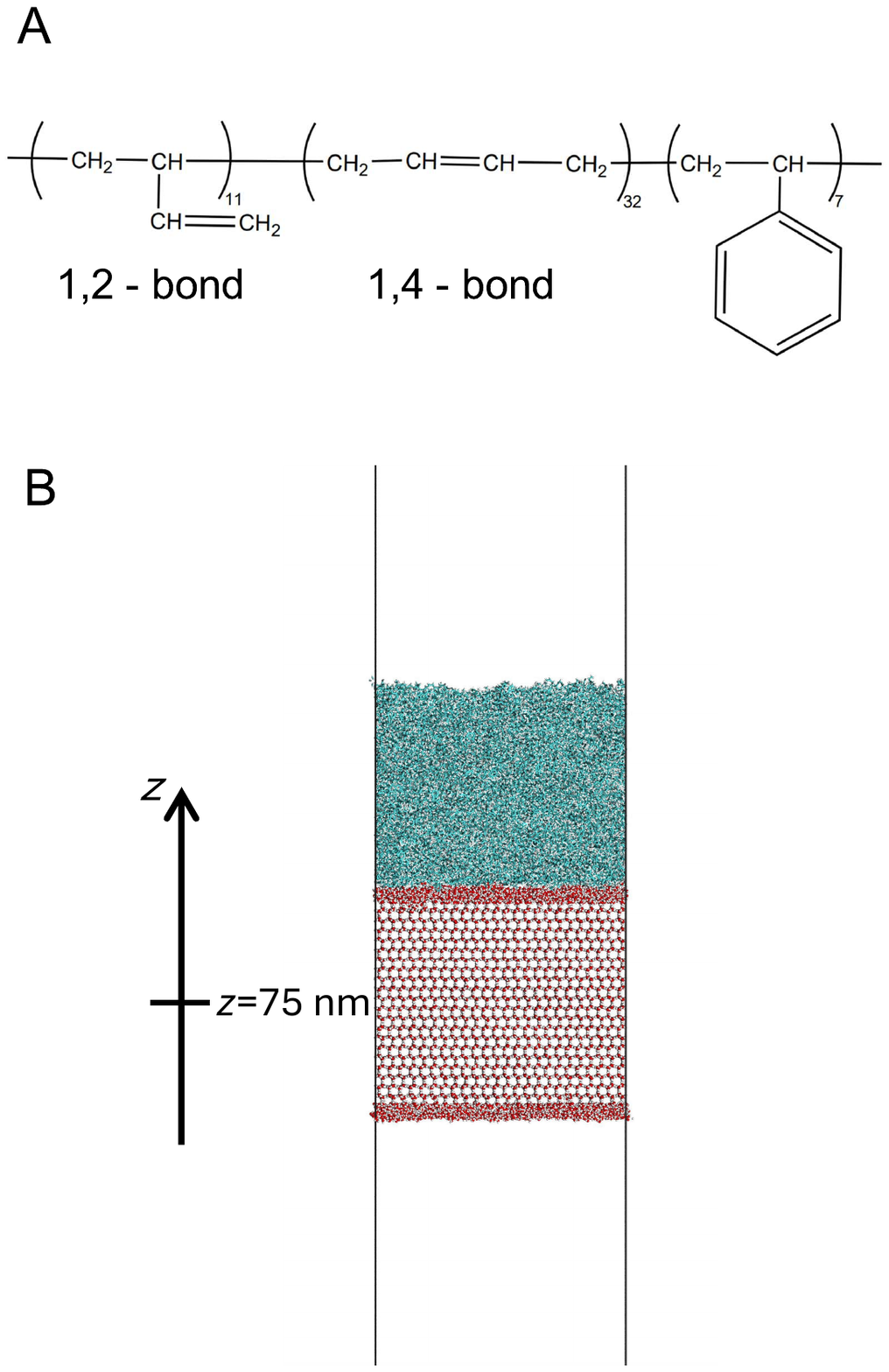}
    \caption{
    Simulation system.
    (A) Chemical structure of styrene-butadiene rubber used in this work. 
    \textcolor{black}{
    We generated a single chain by randomly shuffling the monomers, using the indicated number of monomer units. 
    }
    (B) \textcolor{black}{Representative zoomed-in snapshot of the interface, with the full simulation system shown in Fig.~\ref{fig:snapshot_full_simulation_box}.} The top side is the ice--rubber interface, while the lower is the ice--vapor interface.
    }
    \label{fig:fig1}
\end{figure}

We first created bulk systems of rubber and ice separately. These systems were subsequently combined in a side-by-side arrangement to establish the ice–rubber interface.
The rubber used in this study is styrene-butadiene rubber (SBR) (Fig~\ref{fig:fig1}A), a rubber commonly employed in industrial automobile tires. 
The degree of polymerization of the rubber is 50, composed of styrene, 1,2-butadiene, cis-1,4-butadiene, and trans-1,4-butadiene, mixed at the ratio shown in the Tab.~\ref{tab:tab1}. 
\textcolor{black}{The glass transition temperature of this composition is reported as $242~\mathrm{K}$~\cite{zhang2021quantitatively}.}
The polymer sequence was generated by randomly shuffling the monomer units using the molecular modeling software J-OCTA (\url{http://www.j-octa.com/}). 
We randomly inserted 140 copies of the rubber chain into a large cubic box, which was then scaled to the final simulation box size. To equilibrate the rubber conformations, simulated annealing was performed over three cycles in the temperature range of 254--1000~K.
Ice 1h structure was generated in 12$\times$12$\times$12 unit structures using GenIce~\cite{Matsumoto:2024}. 
$NPT$ equilibration was performed to determine the unit cell dimensions at 0.1~MPa and for each temperature (254, 259, 264, 267, 269~K). This temperature range is below the melting point of TIP4P/Ice, $269.8\pm0.1$~K~\cite{conde2017high}.
Force field parameters were GAFF2~\cite{wang2004development} and TIP4P/Ice~\cite{abascal2005potential} for rubber and ice, respectively. 
The bulk systems of rubber and ice were combined in a side-by-side configuration to form an interface, with the basal face of the ice in contact with the rubber. The slab geometry was then constructed by extending the system along the direction normal to the interface (Fig.~\ref{fig:fig1}B and Fig.\ref{fig:snapshot_full_simulation_box}). This system setup enables the simultaneous simulation of both ice–rubber and ice–vapor interfaces.  
The unit cell geometry was about 9.43~nm$\times$10.9~nm$\times$150~nm across the temperature range.
\textcolor{black}{
Periodic boundary conditions are applied in all directions. 
Although we included a large empty space in our systems to avoid immediate transfer of water molecules between the ice and rubber surfaces through the periodic boundaries, the amount of empty space could be reduced to improve the computational efficiency of particle-mesh Ewald (PME) calculations.
}

\begin{table}[tb]
  \begin{tabular}{lll}
    \hline
    Chemical species & mol~\% & wt~\% \\
    \hline
    Styrene              & 13.8 & 23.3  \\
    1,2-Butadiene       & 22.5 & 20.0  \\
    Cis-1,4 -butadiene  & 10.0 & 8.9   \\
    Trans-1,4 butadiene & 53.7 & 47.8  \\
    \hline
  \end{tabular}
  \caption{SBR composition}
  \label{tab:tab1}
\end{table}

Molecular dynamics simulations were performed in $NVT$ ensemble with $p_z=1~\rm{MPa}$
\textcolor{black}{to match the ice-rubber interfacial conditions reported in previous experiments~\cite{stevens2025viscoelasticity}..} 
\textcolor{black}{Instead of applying pressure coupling along the $z$ axis, we directly applied external forces to pull the ice and rubber atom groups toward each other in opposite directions using the \textit{pull} command in GROMACS.
The magnitude of the constant force, $F$, was calculated from the target pressure using the equation $F=P \cdot S$, where $P$ is the target pressure and $S$ is the surface area of the simulation box.}
\textcolor{black}{Our approach enables the application of an arbitrary interfacial pressure, making it more appropriate for reproducing experimental conditions.}
Smooth PME method~\cite{essmann1995smooth} was used to calculate the electrostatic interactions, using 4th-order interpolation and the Fourier spacing of 0.12~nm. 
The cutoff distance for Van der Waals interactions and the short-range cutoff for PME were both set to 1.13~nm. 
\textcolor{black}{The internal O-H bonds of TIP4P/Ice water molecules were constrained using the SETTLE algorithm~\cite{miyamoto1992settle}.} 
Temperature was controlled using the Nos$\acute{\rm{e}}$-Hoover thermostat~\cite{Nose1984thermostat, Hoover1985thermostat}, with the coupling time of 0.1~ps.
The leapfrog integrator was used with a time step of 1~fs, and trajectories were saved every 1~ps. 
After equilibration for more than 50~ns, production runs continued for more than 50~ns.
All the simulations were performed using GROMACS 2022.4~\cite{abraham2015gromacs}, and trajectories were visualized using VMD~\cite{humphrey1996vmd}.

\subsection{Density Profile}
The density profile was computed based on atomic masses after re-centering the system on the inner bulk region of the ice crystal. A bin width of 0.02~nm was used for the analysis.

\subsection{Diffusion coefficient}
\textcolor{black}{Molecular dynamics of premelting layers was evaluated using the mean square displacement (MSD) along the directions parallel to the interface},
\begin{equation}
    \begin{split}
    \mathrm{MSD}(\Delta) = \langle |\bm{r}(t+\Delta)-\bm{r}(t)|^2 \rangle
    \end{split}
\end{equation}
where $\langle \cdot \rangle$ means the ensemble average, and $\bm{r}(t)$ is the position of the oxygen atom. 
\textcolor{black}{In this calculation, we counted molecules that remained within the specified region for the measurement time, $\Delta$: (1) Layer 1 and (2) the combined region of Layers 1 and 2.} 
A comparison of these definitions is shown in the Results section.  
\textcolor{black}{The parallel diffusion coefficient~\cite{kling2018structure, louden2018ice, baran2022ice}, $D$, was computed from the \textcolor{black}{slope} of MSD in the nearly linear region (after lag time 100~ps),
\begin{equation}
    D = \frac{1}{2d}\frac{d}{dt}\mathrm{MSD}(t)
\end{equation}
where $d$ is the number of dimension and, for the parallel diffusion $d=2$.}

\subsection{Analysis of single molecule dynamics using machine learning}
Although MSD is useful for characterizing system-averaged molecular dynamics, single-molecule dynamics cannot be fully captured. 
\textcolor{black}{Although bond-order parameters have been widely used to detect local ice structures~\cite{pickering2018grand, mochizuki2023microscopic}, these descriptors rely on instantaneous local structures. However, the polymer–ice interfaces are highly heterogeneous, making it difficult to judge the local states based on single-frame structural snapshots.}
Therefore, to distinguish liquid-like mobile molecules from solid-like immobile molecules, a machine-learning approach for single-molecular dynamics analysis has been developed~\cite{endo2019detection, yasuda2023combining, yasuda2024layer}.
In this work, we employed this method to clarify the rubber-induced effect on the premelting molecules at single-molecule resolution. In the machine learning approach, we first sample the short-term trajectories from the rubber-ice interface, $\bm{x}$, and those from bulk ice systems, $\bm{x'}$. Any type of trajectories can be used for $\bm{x'}$, but we employed a time series of the oxygen atom displacement in the water molecules for 64~ps, according to our previous work~\cite{yasuda2024layer}. Both $\bm{x}$ and $\bm{x'}$ are input into a neural network, which outputs a score, $g(\bm{x})$, that quantifies the difference from the solid bulk water system,
\begin{equation} 
    g(\bm{x}) = \mathbb{E}_{\bm{x}' \sim \bm{y}'} \left[ f^*(\bm{x}) - f^*(\bm{x}') \right] 
\end{equation} 
where $\mathbb{E}$ represents the expectation value taken over samples $\bm{x}$ sampled from the distribution of $\bm{y}$, and $f^*(\bm{x}')$ is the function represented by the neural network, which is a simple multi-layer perceptron. The neural network model trained in ref.~\cite{yasuda2024layer} was used. Higher values of $g(\bm{x})$ correspond to molecular dynamics that differ from the solid phase. 
Following our previous work, a threshold value of $g(\bm{x})$, which was determined based on the comparison of liquid and solid bulk systems, was used to distinguish liquid and solid molecules in the premelting layers.

\subsection{Analysis of polymer fluctuations}
We quantified the interfacial fluctuations of rubber by calculating the root mean square fluctuation (RMSF) of non-hydrogen rubber atoms,
\begin{equation}
\mathrm{RMSF} = \langle | \bm{r}(t)- \bar{\bm{r}} | \rangle,
\end{equation}
where $\bm{r}$ is the atom position, and $\bar{\bm{r}}$ is the time-averaged position of each atom. 
Rubber atoms within the range $79.05~\text{nm} \leq z \leq 80.06~\text{nm}$, which includes both the premelting–mixing layers and part of the bulk interfacial rubber, were used consistently across all temperatures.
To evaluate short-term dynamics, the trajectory was segmented into 5~ns windows, and the RMSF was computed separately for each segment. Trajectories every 100~ps were used in this computation to remove highly correlated frames.

\section{Results and Discussion}
\subsection{Enhanced disorder of premelting layers at the ice--rubber interface}
\begin{figure*}[tbh!]
    \includegraphics[width=120mm]{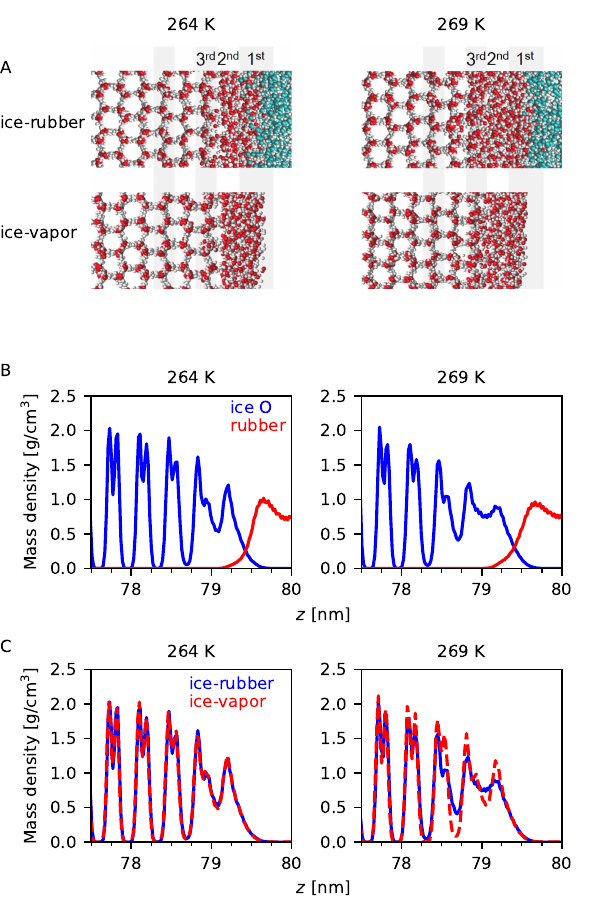}
    \caption{
    Molecular structure of ice--rubber and ice--vapor interfaces.
    (A) Representative snapshots at 264 and 269~K for the ice--rubber and ice--vapor interfaces. Labels (1st, 2nd, and 3rd) indicate the layer positions (Layer 1--3). 
    (B) Density profile of water oxygen atoms and rubber atoms in the ice--rubber interface at 264 and 269~K.
    (C) The density profile of water oxygen atoms, comparing the ice--rubber and ice--vapor interfaces, at 264 and 269~K.
    }
    \label{fig:fig2}
\end{figure*}

As premelting layers are highly sensitive to temperature near the melting point~\cite{sazaki2012quasi,sanchez2017experimental}, we compare the atomistic surface structures at 264 and 269~K for the ice--vapor and ice--rubber interfaces.
Snapshots show that the surface structures are similar at 264~K, whereas at 269~K, the third layer becomes more disordered at the rubber interface, compared to the vapor interface (Fig.~\ref{fig:fig2}A).
The density profiles show that rubber and premelting layers coexist in a very thin surface layer at both temperatures (Fig.~\ref{fig:fig2}B). 
However, the density profiles at 269~K show weaker peaks at the rubber interface compared to the vapor interface, while their profiles at 264~K are nearly identical (Fig.~\ref{fig:fig2}C). This difference is evident even in the 3rd ice layer, which is not directly exposed to rubber atoms. 
Interestingly, at 269~K, the third layer at the ice--rubber interface exhibits comparable peak heights and widths in the density profile to the second layer at the ice--vapor interface, indicating structural similarity between the two layers (Fig.~\ref{fig:fig2}C).

\textcolor{black}{Comparison of the profiles at different temperatures in the range of 254--269~K (Fig.~\ref{fig:figS1}) indicates a sharp increase in ice premelting between 264 and 267~K}, accompanied by enhanced penetration of rubber atoms into the premelting layer.
\textcolor{black}{
To further support the disruption of the hydrogen-bond network, we computed the orientational order parameter (Fig.~\ref{fig:order_parameter}). The reduction in peak intensity indicates rubber-induced orientational disordering above 264 K in Layers 2--4 (Fig.~\ref{fig:order_parameter_peak}), in consistent with the density profiles.
}

\subsection{Suppressed water dynamics by rubber confinement}
\begin{figure*}[tb!]
    \includegraphics[width=160mm]{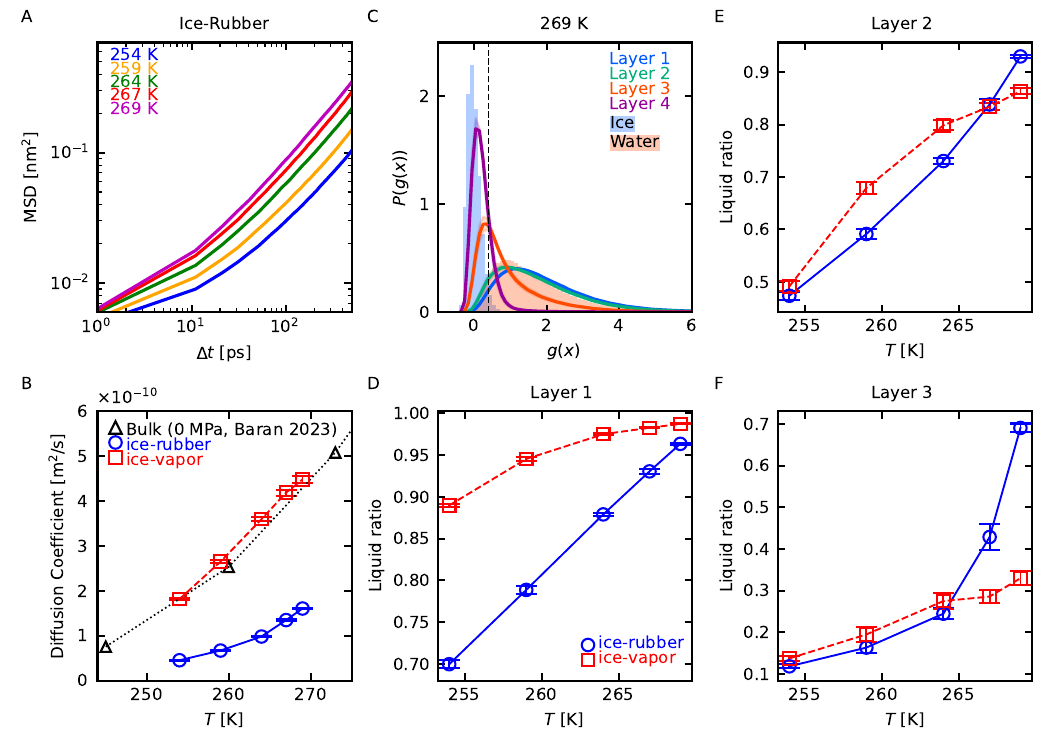}
    \caption{
    Comparison of the dynamical properties of ice--rubber and ice--vapor interfaces.
    (A) Mean squared displacement (MSD) of premelting layers (combined Layer 1 and Layer 2) in the ice–rubber interface.
    $\alpha$ is the scaling exponent of MSD.
    (B) \textcolor{black}{Parallel diffusion coefficient as a function of temperature.
    Bulk liquid values at 0 MPa were taken from Baran et al.~\cite{baran2023self}.}
    Values of the parallel diffusion coefficients are listed in Tab.~\ref{tab:tabS1_combined}.
    (C) Distribution of the score, $g(\bm{x})$, which quantifies deviation from bulk solid behavior in single-molecule short-term dynamics. 
    Layer 1--4 represent the distributions from the ice–rubber interface at 269~K, while "Ice" and "Water" correspond to bulk solid and liquid systems, respectively. 
    The black dashed line indicates the threshold used to classify molecules as solid-like or liquid-like.
    (D–F) Ratio of liquid-like molecules in Layers 1--3 at different temperatures. 
    Error bars represent the standard error of the mean calculated using time block averaging (5 ns per block) in (B) and (C--F).  
    }
    \label{fig:fig3}
\end{figure*}

Building on the observation of enhanced structural disorder at the ice--rubber interface, we use dynamical properties to clarify whether the premelting layers behave as mobile liquid.
\textcolor{black}{MSD was computed for water molecules that remained in Layer 1 at the ice--vapor and at ice--rubber interfaces.}
For the rubber interface, MSD increased significantly between 264 and 269~K (Fig.~\ref{fig:fig3}A). 
\textcolor{black}{The scaling exponent of MSD, $\alpha$ ($\rm{MSD} \propto t^{\alpha}$), is 1 for Brownian motion, and if the exponent is lower than 1, the dynamics are subdiffusive, which is typically observed in glassy or viscoelastic materials and crowding environments~\cite{kling2018structure}.}
The scaling exponent ranged in 0.92--0.96 for the vapor interface, and 0.77--0.85 for the rubber interface (Tab.~S1), indicating the glassy behavior of the premelting layers \cite{kling2018structure}.
The ice--rubber interface showed a lower $\alpha$ value compared to the ice--vapor interface, suggesting that the presence of rubber molecules enhanced the confinement of water molecules within the premelting layers.

To further clarify the temperature dependence of interfacial dynamics, we computed the parallel diffusion coefficients for the ice--vapor and ice--rubber interfaces. As shown in Fig.~\ref{fig:fig3}B, the parallel diffusion coefficient at the ice--vapor interface increased linearly with temperature, whereas the ice--rubber interface exhibited a change in the \textcolor{black}{slope} between 264 and 267~K. 
These results highlight distinct differences in both diffusivity and temperature dependence between the two interfaces.
\textcolor{black}{Furthermore, our results demonstrate higher the parallel diffusion coefficients at the ice--vapor interface than the diffusion coefficient of bulk liquid, consistent with previous studies~\cite{kling2018structure, louden2018ice}, whereas the rubber interface exhibited suppressed dynamics (Fig.~\ref{fig:fig3}B). In contrast, a previous study reported enhanced dynamics at smooth hydrophobic interfaces~\cite{baran2022ice}. The suppressed dynamics in our atomistic rubber interfaces showed suggests that the structural roughness of the interface induced confinement effects.
}   

\textcolor{black}{    
We compared the parallel diffusion coefficients at the rubber interface for (1) Layer 1 alone and (2) the combined Layers 1 and 2. At most temperatures, Layer 1 alone showed higher diffusion coefficients than the combined Layers 1 and 2. However, at 269 K, the combined Layers 1 and 2 showed a higher diffusion coefficient. We interpret this as the result of polymer-induced confinement within Layer 1, while Layer 2 provides a partially melted ice structure that facilitates water diffusion (Fig.~\ref{fig:MSD_combined_layer1_2}).  
}

\textcolor{black}{While these simulations were performed at $1~\mathrm{MPa}$, the density profiles at \textcolor{black}{254} and 269~K are nearly the same at $0.1~\mathrm{MPa}$, indicating that the pressure effect is negligible within this range (Fig.~\ref{fig:density_profiles_01_1MPa_254_269}).
Similarly, the difference in the diffusion coefficient between 0.1 MPa and 1 MPa was minor (Fig.~\ref{fig:1MPa_01MPa_MSD}).}

\subsection{Liquid-like molecular ratio for each layer}
Besides the mean behavior within the layer captured by MSD, we speculate that the local nanostructure of the rubber molecules contributes to the heterogeneity in the layer, thus requiring a single-molecular dynamics analysis.
Here, from the measurement of the short-term molecular dynamics (64~ps) of ice premelting layers at the single-molecule resolution~\cite{yasuda2024layer}, we quantify the ratio of mobile-liquid-like molecules.
\textcolor{black}{
This approach assigns to each short-term single-molecule trajectory, $\bm{x}$, a score, $g(\bm{x})$, that quantifies how different its dynamics are from those observed in the bulk ice. The model was trained on bulk solid and bulk liquid water at 269~K.
Despite being trained only on bulk systems at a single temperature, the model successfully generalizes to interface systems, including different ice crystal faces across different temperatures~\cite{yasuda2024layer}. In our ice--rubber interfaces, this is evidenced by the gradual shifts in $g(\bm{x})$ distributions with temperature (Fig.~\ref{fig:gx_profiles_254_267}), demonstrating the robustness of the approach. 
Focusing on 269~K, we computed the score $g(\bm{x})$ for the four premelting layers, and compared the score to those of bulk solid and liquid water systems. 
For the rubber interface, Layers 1 and 2 show similar properties to the bulk liquid system (Fig.~\ref{fig:fig3}C), indicating that these layers predominantly consist of liquid-like molecules.
}

\textcolor{black}{
The threshold for distinguishing liquid-like from solid-like molecules was optimized at 269~K to achieve balanced classification accuracy for both solid and liquid molecules~\cite{yasuda2024layer}.
We applied this threshold consistently across all temperatures. 
To validate the threshold at the ice--rubber and ice--vapor interfaces, we examined Layer 4, which serves as a reference region consisting predominantly of solid-like molecules. 
Specifically, the liquid ratio in Layer 4 was about 6\% at the ice-vapor interface, and 8\% at the ice-rubber interface at 254~K (Fig.~\ref{fig:Liquid_Ratio_Layer4}), which are sufficiently low values. At the higher temperatures, it increased to about 15\% at the ice-vapor interface, which is the similar liquid ratio as obtained in our previous work~\cite{yasuda2024layer}. 
The difference in the liquid ratio between the bulk solid and Layer 4 at the ice--vapor interfaces at 269 K is attributed to both the uncertainty of the model and interface-enhanced mobility. Considering these points, the threshold reasonably captures the differences between solid--like and liquid-like dynamics, as a semi-qualitative measurement.
}


We compared the rubber and vapor interfaces in terms of the ratio of the mobile-liquid-like molecules. 
In the topmost layer of the rubber interface, the ratio of liquid-like molecules linearly increases, while that in the vapor interface shows an overall higher liquid-like ratio, yet saturating at above 259~K (Fig.~\ref{fig:fig3}D). 
In Layer~2, the rubber interface shows lower liquid-like ratios than the vapor interface below 267~K (Fig.~\ref{fig:fig3}E). However, this relationship switched at 269~K, with the liquid-like ratios of the rubber interface approaching those in Layer~1. 
This trend is consistent with the structural disordering (Fig.~\ref{fig:fig2}A and C).
Layer~3 shows a similar trend to Layer~2, showing an unexpectedly high liquid-like ratio at the rubber interface (Fig.~\ref{fig:fig3}F).
These results show that, at the rubber interface, molecular dynamics are restrained due to the presence of rubber molecules. 
However, near the melting point, the topmost layer mixes with the rubber, causing the bottom layers to locate near the interface and allowing them to behave similarly to the previous topmost layers.
Despite the similarity in density profiles at 269~K between Layer 3 at the rubber interface and Layer 2 at the vapor interface (Fig.~\ref{fig:fig2}C), the liquid ratio was approximately 0.7 for the former and nearly 0.9 for the latter, indicating comparable but distinct molecular dynamics in the two layers.

\begin{figure*}[tb!]
    \includegraphics[width=140mm]{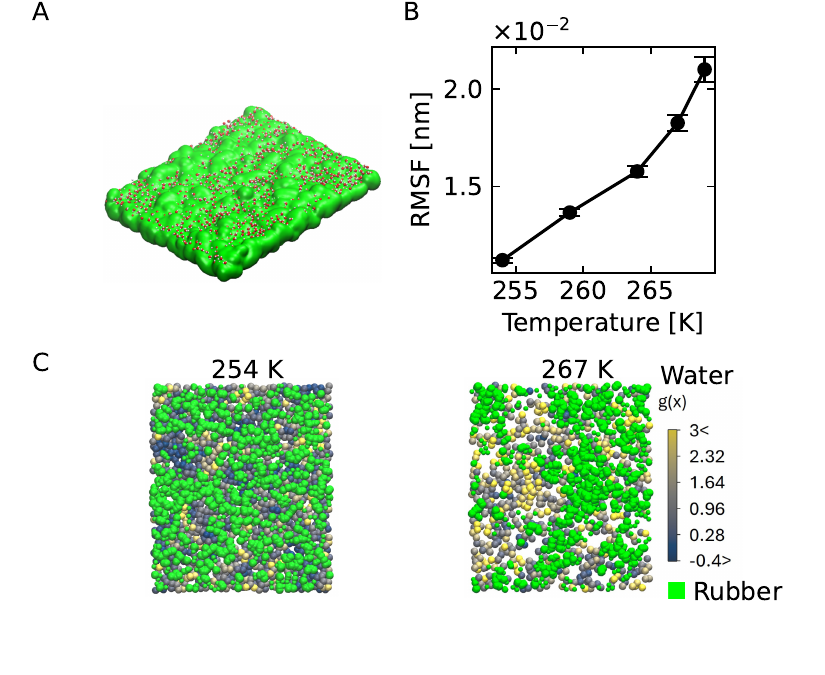}
    \caption{
    Analysis of the atomistic structure of the ice–rubber interface.
    (A) Snapshot of Layer 1 at 269~K, showing water molecules as Van der Waals spheres and rubber atoms as a surface representation.
    (B) Root-mean-square fluctuation (RMSF) of surface rubber atoms in Layer 1. Error bars represent the standard error of the mean calculated using time block averaging (5~ns per block).
    (C) Representative snapshot of a region within Layer 1, where the densities of water and rubber are comparable, at 254 and 267~K. Water molecules are shown as oxygen atom positions and are colored by the score $g(\bm{x})$, which quantifies deviation from bulk ice. Higher values (yellow) indicate greater mobility. Rubber atoms are shown as green spheres.
    }
    \label{fig:fig4}
\end{figure*}

\subsection{Coupling of the dynamics of rubber molecules and premelting layers}
We have shown the rubber--induced effects on ice premelting layers, which are attributed to the molecular interactions and nanoscale surface structure of the rubber. 
We therefore conducted a detailed analysis of the atomistic interface of the rubber.
The snapshot (Fig.~\ref{fig:fig4}A) shows that, despite the hydrophobic nature of the rubber chains, water molecules penetrate the rubber matrix. 

We hypothesized that, as the ice--rubber interface is tightly packed, the dynamics of rubber and ice molecules become coupled. To test this, we computed the RMSF of the rubber atoms within Layer~1 (Fig.~\ref{fig:fig4}B). The fluctuations increased steeply with temperature rise, showing a marked rise above 264~K, consistent with the temperature dependence of water dynamics at the ice--rubber interface (Fig.~\ref{fig:fig3}B).
Furthermore, the spatial distributions of rubber and water molecules within Layer 1 indicate that the packing was looser at 267~K than at 254~K (Fig.~\ref{fig:fig4}C), leading to increased rubber mobility.
Nevertheless, rubber atoms, mobile water molecules, and immobile water molecules each formed distinct clusters within the layer at both temperatures.

\section{Conclusions}
Tribology of ice–rubber interface is influenced by multiple factors, including frictional heating, pressure melting, ice adhesion, and premelting layers, yet a comprehensive understanding of these effects remains elusive at atomistic resolution. 
\textcolor{black}{
In previous studies on ice or liquid–water interfaces, idealized, atomically smooth hydrophobic walls were employed, revealing depletion effects and enhanced water mobility~\cite{janecek2007interfacial, baran2022ice}. 
In contrast, recent studies have modeled ice interfaces using fully atomistic surface representations of materials such as graphene and polymers, showing that atomistically heterogeneous surfaces enable site-specific interactions between water and particular surface atoms~\cite{janecek2007interfacial, sato2025molecular}, but such effects cannot be captured by smooth-wall models.
However, no prior molecular simulation has examined ice in contact with an atomistic rubber surface models, leaving the molecular origins of rubber-induced premelting entirely unexplored.}

\textcolor{black}{
In this study, we employed an atomistic rubber model for the interface, demonstrating that rubber, despite its hydrophobicity, induces structural disorder at the ice interface, thereby enhancing ice premelting.
In contrast, based on the diffusion coefficient and liquid-like ratio, the premelting layer at the ice--rubber interface shows lower mobility compared to that at the ice--vapor interface. 
Notably, at temperatures approaching the melting point (267 to 269~K), a drastic increase in molecular dynamics was observed at the rubber-ice interface, accompanied by enhanced rubber molecular dynamics. 
These results demonstrate that ice and rubber mutually influence each other's structure and dynamics, resulting in an ice-rubber interfacial layer with unique temperature-dependent properties.
}

Why do hydrophobic polymers enhance premelting? Previous studies have proposed a depletion mechanism \cite{janecek2007interfacial, poynor2006water, chen2017icephobic}, in which water molecules are excluded from the hydrophobic polymer surface due to repulsive excluded volume effects, forming a water depletion layer.
In contrast, our results show that water molecules can penetrate nanoscale voids within the polymer structures (\textcolor{black}{Fig.~\ref{fig:fig4}a}), where the thickness of roughness is comparable to the bilayer thickness of ice. This suggests that the nanoscale morphology of the polymer, via the excluded volume effect, disrupts hydrogen-bond networks of the water molecules within the premelting layers (\textcolor{black}{Fig.~\ref{fig:order_parameter}})~\cite{kimmel2009no}.
Notably, the substantial increase in rubber flexibility observed in our study is likely triggered by bilayer melting, which occurs a few kelvins below the melting point at the ice-vapor interface~\cite{sanchez2017experimental}.
As a result, the topmost layer becomes a mixed rubber--ice state, encouraging lower layers (Layer 2 and Layer 3) to behave as if they are the topmost interface layer.

\textcolor{black}{
Our observations of molecular structure and dynamics can be connected to experimental results.
Surface force apparatus measurements have shown a sharp decrease in viscosity of the premelting layers above $-5~ \mathrm{^\circ C}$~\cite{lecadre2020ice}. 
Because the viscosity and diffusivity are inversely related thorough the Stokes-Einstein equation~\cite{baran2022ice}, these experimental measurements correspond to the increase in the diffusion coefficient of the premelting layers in our simulations. 
In contrast, the apparent contact area of the hydrophobic polymer with ice changed little in experiment~\cite{pallbo2024enhanced}, although we observed increased structural disordering.
This suggests that the increased disordering of premelting layers at the rubber interface observed in our simulations are confined to only a few molecular layers, making them detectable at the molecular scale but likely averaged out at the spatial resolution of the experimental measurements.}

\textcolor{black}{Our simulation setup employed a minimal model, consisting of a single SBR rubber composition in contact with the basal face of ice. In practical applications, however, rubber formulations typically vary in composition and contain cross-links, while multiple ice crystal faces may be exposed depending on environmental conditions.
Our study serves as a baseline system that clarifies key physical phenomena, such as confined premelting dynamics and the coupled interfacial dynamics between ice and rubber chains. This framework can be systematically extended in future studies to incorporate additional material complexities, including fillers, copolymer architectures, surface treatments, and different ice crystal faces.}

Ice–rubber tribology is a multiscale phenomenon, and our work provides fundamental molecular-level insights into the rubber–ice interface.
\textcolor{black}{Although we have focused on the equilibrium properties, in real-life conditions, the ice interface is often subjected to sliding, and under such conditions, the effective viscosity of the liquid decreases due to shear thinning~\cite{zhao2022new, baran2022ice, baran2024confinement}.}
We speculate that the premelting layers influence both the effective friction coefficient and adhesion to ice. Bridging these molecular-level understanding with macroscopic observations may contribute to the development of functional polymer materials and their industrial applications.

\section*{Supporting Information}
Method for calculating orientational order parameters; a table of diffusion coefficients and scaling exponents; additional figures including a snapshot of the full simulation box, density profiles, orientational order parameters, mean square displacements, diffusion coefficients, $g(x)$ distributions, and liquid-like ratio.


\section*{Acknowledgments}
This work was supported by JST, CREST (Grant No. JPMJCR2093), and in part by MEXT under the “Program for Promoting Research on the Supercomputer Fugaku” (Grant No. JPMXP1020230325). The authors are grateful to Prof. K. Kurihara (Tohoku University) and Prof. M. Mizukami (Tohoku University) for helpful discussions and valuable comments. I.Y. was supported by a Grant-in-Aid for JSPS Fellows (Grant No. JP23KJ1918).

\bibliography{references}

\newpage
\section*{Supporting Information}

\setcounter{figure}{0}
\renewcommand{\thefigure}{S\arabic{figure}}
\setcounter{table}{0}
\renewcommand{\thetable}{S\arabic{table}}

\subsection{Orientation order parameter}
To quantify the structural ordering of the ice premelting layer, we computed the orientational order parameter and the intensity of the peak \cite{sanchez2017experimental}. The orientational order parameter was defined as,
\begin{equation}
    \cos{\theta} = \frac{\bm{n}_{z} \cdot \bm{r}_{\rm{OH}} }{|\bm{r}_{\rm{OH}}|}
\end{equation}
where $\bm{n}_{z}$, $\bm{r}_{\rm{OH}}$ are the normalized vector in the surface direction and the vector from the oxygen atom to the dummy charge. 


\begin{table}[htbp]
    \centering
    \renewcommand{\arraystretch}{1.5} 
    \begin{tabular}{|c|c|c|c|c|}
        \hline
        \textbf{Temperature} & \multicolumn{2}{c|}{\textbf{Vapor}} & \multicolumn{2}{c|}{\textbf{Rubber}} \\
        \cline{2-5}
        \textbf{[K]} & \textbf{$ \alpha $} & \textbf{Diffusion Coefficient [$\mathrm{m^2/s}$]} & \textbf{$ \alpha $} & \textbf{Diffusion Coefficient [$\mathrm{m^2/s}$]} \\
        \hline
        254 & 0.922 & $ 1.817 \times 10^{-10} $ & 0.778 & $ 4.548 \times 10^{-11} $ \\
        \hline
        259 & 0.949 & $ 2.649 \times 10^{-10} $ & 0.808 & $ 6.704 \times 10^{-11} $ \\
        \hline
        264 & 0.953 & $ 3.596 \times 10^{-10} $ & 0.825 & $ 9.828 \times 10^{-11} $ \\
        \hline
        267 & 0.958 & $ 4.187 \times 10^{-10} $ & 0.860 & $ 1.349 \times 10^{-10} $ \\
        \hline
        269 & 0.963 & $ 4.480 \times 10^{-10} $ & 0.854 & $ 1.606 \times 10^{-10} $ \\
        \hline
    \end{tabular}
    \caption{Diffusion exponent~(2D, Layer1), $\alpha$ ($\rm MSD \propto t^\alpha$), and diffusion coefficient (2D, Layer1) at the ice--vapor interface (Vapor) and the ice--rubber interface (Rubber).}
    \label{tab:tabS1_combined}
\end{table}

\begin{figure*}[h!]
    \includegraphics[width=150mm]{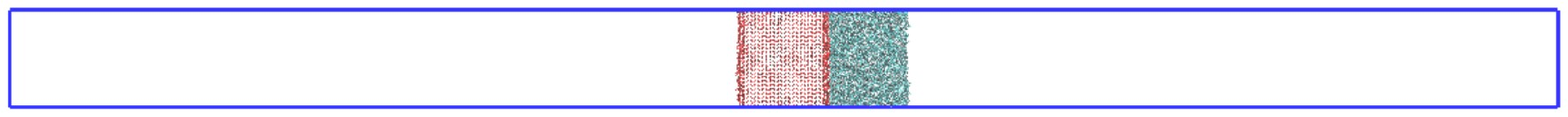}
    \caption{
    \textcolor{black}{
    Snapshot showing the full simulation box. 
    }
    }
    \label{fig:snapshot_full_simulation_box}
\end{figure*}

\begin{figure*}[h!]
    \includegraphics[width=150mm]{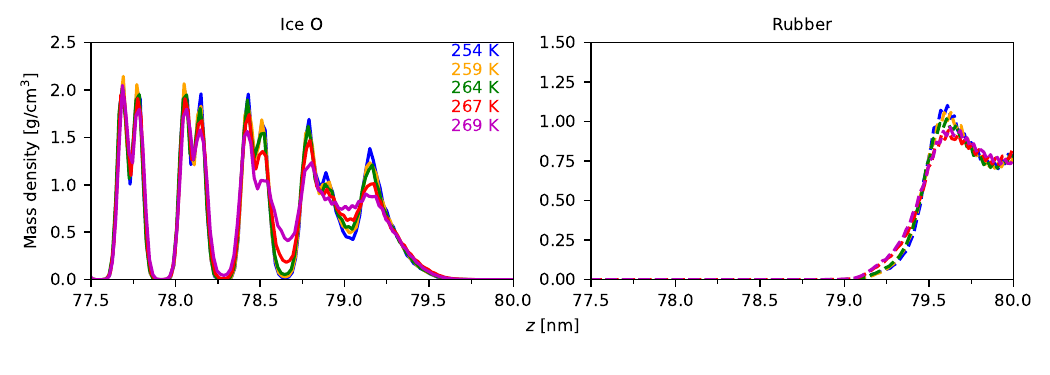}
    \caption{
    Density profiles of water oxygen atom (Ice O) and rubber atoms (Rubber) at 254, 259, 264, 267 and 269~K.
    }
    \label{fig:figS1}
\end{figure*}

\begin{figure*}[h!]
    \includegraphics[width=100mm]{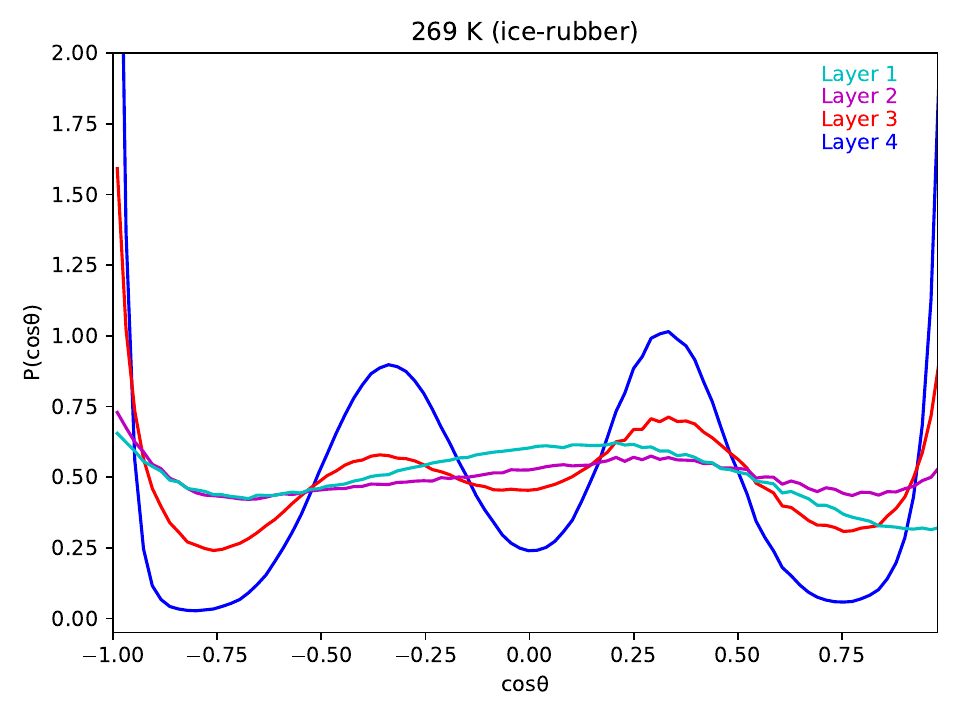}
    \caption{
    \textcolor{black}{
    Probability density function of the orientational order parameter, $P(\theta)$, defined as the angle, $\theta$, between the dipole moment and the interface normal direction.
    }
    }
    \label{fig:order_parameter}
\end{figure*}

\begin{figure*}[h!]
    \includegraphics[width=150mm]{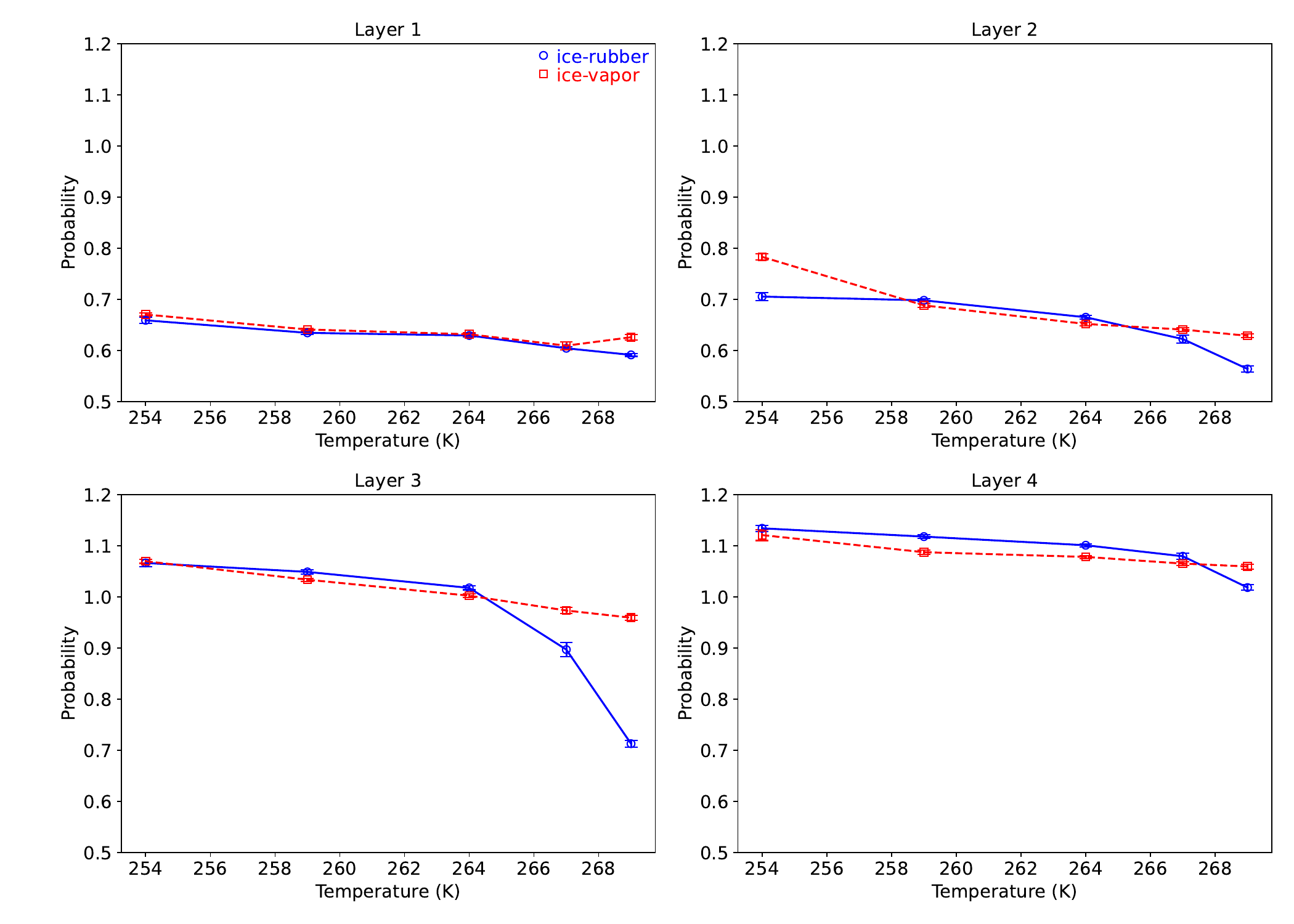}
    \caption{
    \textcolor{black}{Peak values of the probability density function for the orientational order parameter at $\cos\theta \approx 0.31$, where the peak position was determined from Layer 4 at each temperature.}
    }
    \label{fig:order_parameter_peak}
\end{figure*}

\begin{figure*}[h!]
    \includegraphics[width=120mm]{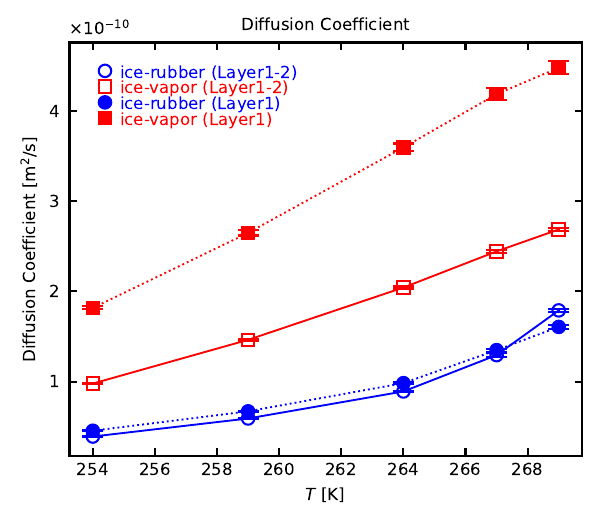}
    \caption{
    \textcolor{black}{Parallel diffusion coefficient calculated for molecules located in Layer 1 (Layer 1) and in the combined Layers 1 and 2 (Layer 1--2) at both ice--vapor and ice--rubber interfaces.}
    }
    \label{fig:MSD_combined_layer1_2}
\end{figure*}

\begin{figure*}[h!]
    \includegraphics[width=150mm]{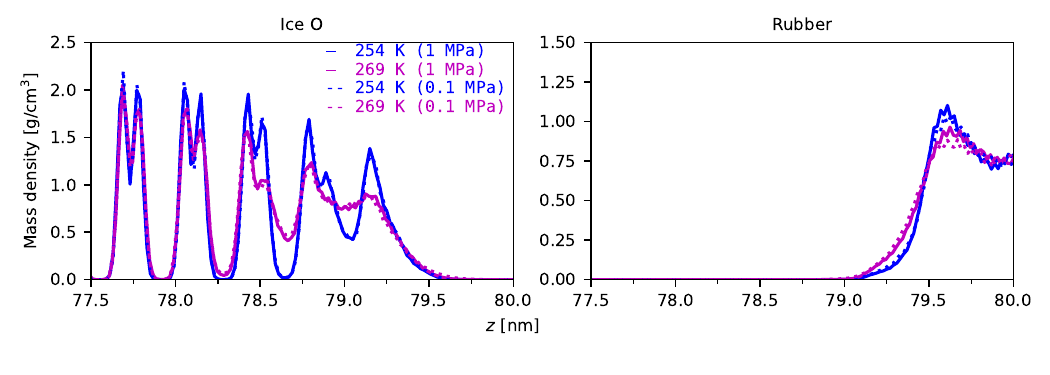}
    \caption{
    \textcolor{black}{
    Density profiles at the ice–rubber interface under load pressures of $0.1~\mathrm{MPa}$ and $1~\mathrm{MPa}$ at 254 and 269~K.
    }
    }
    \label{fig:density_profiles_01_1MPa_254_269}
\end{figure*}

\begin{figure*}[h!]
    \includegraphics[width=150mm]{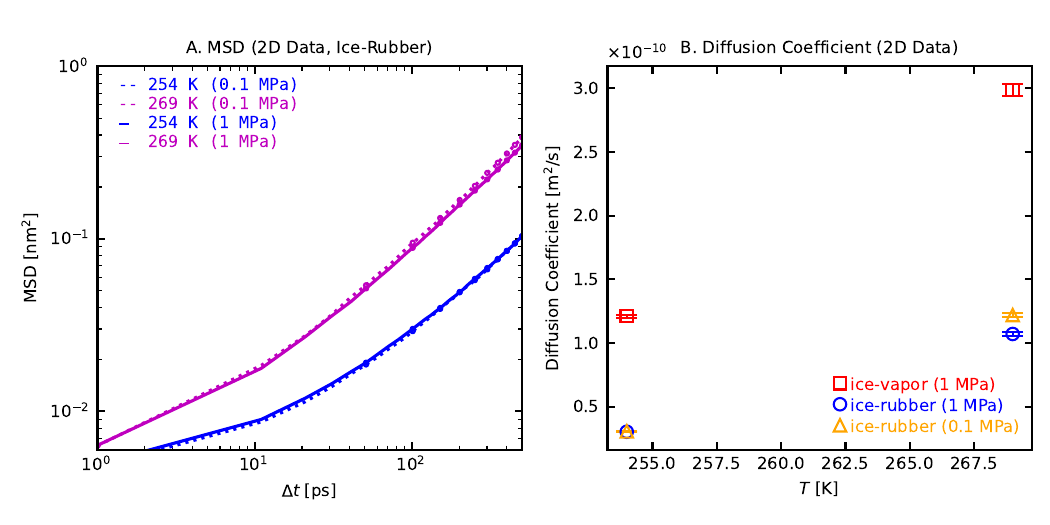}
    \caption{
    \textcolor{black}{
    Diffusive molecular dynamics at $0.1~\mathrm{MPa}$ in comparison with $1~\mathrm{MPa}$. 
    (a) MSD of parallel diffusion. 
    (b) Parallel diffusion coefficient. Simulations at the rubber--ice interface under a load pressure $1~\mathrm{MPa}$ yielded ice--vapor ($0.1~\mathrm{MPa}$) and ice--rubber ($1~\mathrm{MPa}$), while that under a load pressure of $0.1~\mathrm{MPa}$ yielded ice–rubber ($0.1~\mathrm{MPa}$).
    }
    }
    \label{fig:1MPa_01MPa_MSD}
\end{figure*}

\begin{figure*}[h!]
    \includegraphics[width=150mm]{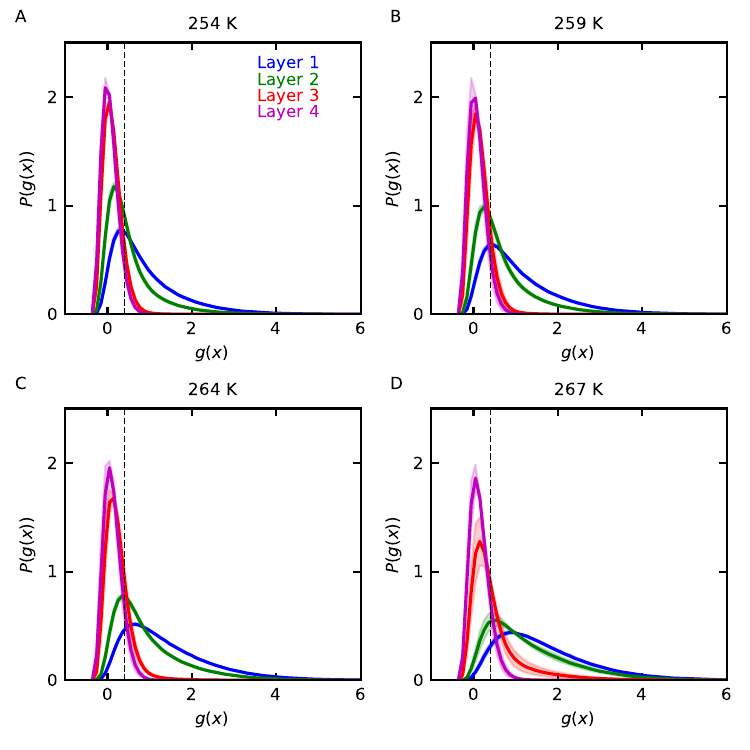}
    \caption{
    \textcolor{black}{
    Profiles of $g(x)$ for ice--rubber interfaces at 254, 259, 264, and 267 K. Dashed black lines indicate the threshold used to distinguish solid-like and liquid-like molecules.    
    }
    }
    \label{fig:gx_profiles_254_267}
\end{figure*}

\begin{figure*}[t!]
    \includegraphics[width=100mm]{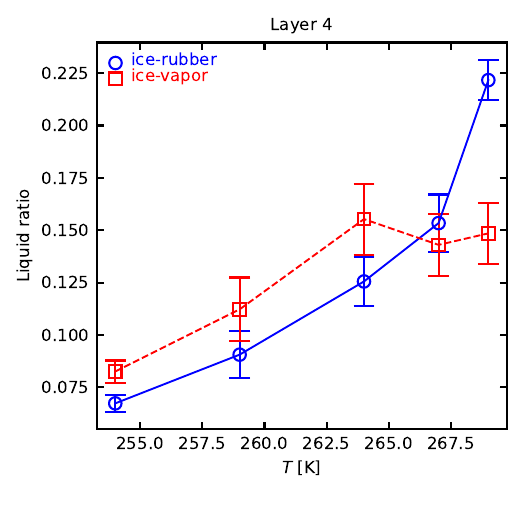}
    \caption{
    \textcolor{black}{
    Liquid ratio in Layer 4 based on $g(x)$. 
    Error bars show the standard error of the mean. 
    }
    }
    \label{fig:Liquid_Ratio_Layer4}
\end{figure*}


\clearpage
\begin{tocentry}
\includegraphics{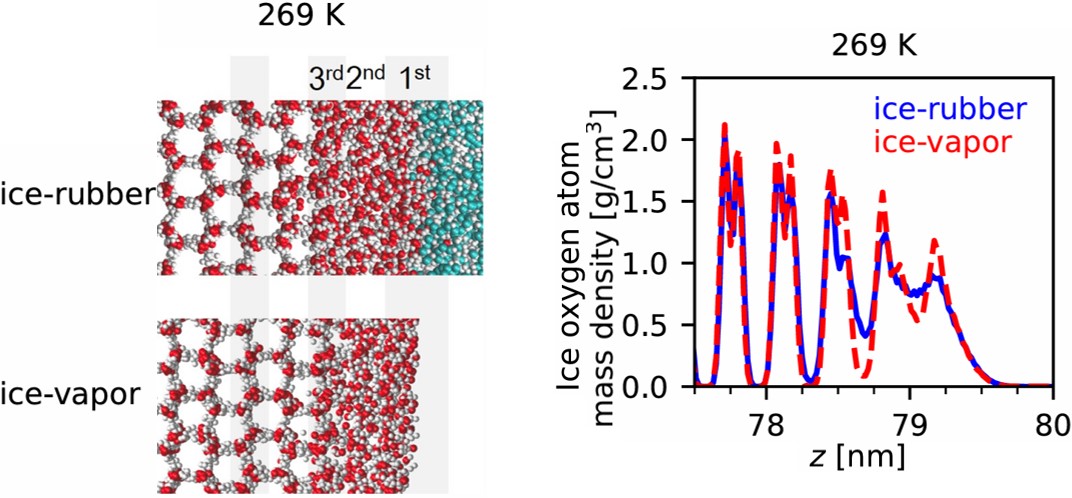}
\end{tocentry}

\end{document}